\documentclass[aps,prl,showpacs,amssymb,nofootinbib,superscriptaddress,twocolumn]{revtex4-1}
\usepackage{tikz}
\usetikzlibrary{shapes,snakes,backgrounds,fit,decorations.pathreplacing}
\usepackage{bbm,times}
\usepackage{bm}
\usepackage{amsbsy}
\usepackage{amsthm}
\usepackage{amssymb}
\usepackage{amsfonts}
\usepackage{amsmath}
\usepackage[c]{esvect} %for nice vectors. Command for vectors becomes \vec{}
\usepackage{dsfont} % for symbols like the identity: \mathds{1}\right 
\usepackage{graphicx} % for graphics
\usepackage{epsfig}
\usepackage{epstopdf}
\usepackage{dsfont}
\usepackage{multibib}
\usepackage{color}

\usepackage{enumerate}
\usepackage{multirow}

\usepackage{braket}

\usepackage[colorlinks]{hyperref}
\makeatletter
\newcommand\org@hypertarget{}
\let\org@hypertarget\hypertarget
\renewcommand\hypertarget[2]{%
	\Hy@raisedlink{\org@hypertarget{#1}{}}#2%
}
\makeatother
\usepackage[figure,table]{hypcap}
\usepackage{MnSymbol}
\usepackage{enumerate}%allows different styles of enumerate environment
\usepackage{float}
\hypersetup{
	bookmarksnumbered,   
	pdfstartview={FitH},
	citecolor={darkblue},
	linkcolor={darkred},
	urlcolor={darkblue},
	pdfpagemode={UseOutlines}}
\definecolor{darkgreen}{RGB}{50,190,50}
\definecolor{darkblue}{RGB}{0,0,190}
\definecolor{darkred}{RGB}{238,0,0}
\usepackage{soul}

\newcommand*\xoverline[2][0.75]{%
    \sbox{\myboxA}{$\m@th#2$}%
    \setbox\myboxB\null% Phantom box
    \ht\myboxB=\ht\myboxA%
    \dp\myboxB=\dp\myboxA%
    \wd\myboxB=#1\wd\myboxA% Scale phantom
    \sbox\myboxB{$\m@th\overline{\copy\myboxB}$}%  Overlined phantom
    \setlength\mylenA{\the\wd\myboxA}%   calc width diff
    \addtolength\mylenA{-\the\wd\myboxB}%
    \ifdim\wd\myboxB<\wd\myboxA%
       \rlap{\hskip 0.5\mylenA\usebox\myboxB}{\usebox\myboxA}%
    \else
        \hskip -0.5\mylenA\rlap{\usebox\myboxA}{\hskip 0.5\mylenA\usebox\myboxB}%
    \fi}
\makeatother

\usepackage[usenames,dvipsnames]{xcolor}

\makeatletter
\renewcommand{\p@subsection}{}
\renewcommand{\p@subsubsection}{}
\makeatother
\begin{document}

\title{Nonclassicality Analysis and Entanglement Witnessing in Spin-$1/2$ NMR Systems} 

\author{Fatemeh Khashami}

\affiliation{Advanced Imaging Research Center, 
University of Texas Southwestern Medical Center, Dallas, Texas, USA\\
\textit{Email: fatemeh.khashami@utsouthwestern.edu}}

\date{\today}

\date{\today}

\begin{abstract}
We investigate quantum features and non-classical nature of two-spin-$1/2$ NMR systems at thermal equilibrium under external magnetic fields. More specifically, using suitable quantifiers, we analyze quantum coherence, mixedness, and entanglement in NMR systems and examine their features within the system. We derive closed-form analytical expressions for the quantum elements and show how they depend on temperature and magnetic field strength. We demonstrate that at zero temperature, the system exhibits a quantum critical point, characterized by non-analytic behavior in the measures of coherence, and a sharp peak in mixedness. Moreover, we analyze the entanglement of the system using a suitable entanglement witness. This provides an experimentally friendly setting for testing entanglement in NMR systems. In other words, the witness links the entanglement in the system to quantum observables, making it directly provable in NMR experiments. We establish a connection between quantum information quantifiers and experimentally accessible NMR spectra of the system, enabling the quantification of entanglement, coherence, and mixedness through NMR signal processing. 
\end{abstract}

\pacs{}
\maketitle

\section{Introduction}

The interaction of two spins in thermal equilibrium under an external magnetic field exhibits a rich quantum behavior, which forms the foundation of phenomena observed in nuclear magnetic resonance (NMR) spectroscopy~\cite{slichter2013principles,bennett1992communication}. In the case of two coupled spin-$1/2$ particles, the system possesses four distinct energy levels. These levels give rise to four possible transitions in the NMR spectrum, capturing the degeneracy and quantum structure inherent to the two-spin system~\cite{ernst1987principles,wootters1998entanglement}.

In two-spin-${1}/{2}$ systems, the interplay between quantum entanglement and magnetic interactions has been extensively studied in different settings~\cite{maleki2021naturalennett,tommasini1997hydrogen,harilal2020hyperfine}. These investigations highlight how entanglement is controlled and modulated by the underlying magnetic coupling as well as temperature. The presence of entanglement in spin systems has attracted significant interest for its potential in quantum information processing and communication~\cite{furman2009nuclear,yamamoto2007feedback,satoori2022entanglement,maleki2015entanglement,broekhoven2024protocol}.

The relationship between thermal polarization and quantum correlations in spin systems remains a rich source of insight for both foundational studies and emerging quantum technologies. Thermal polarization, governed by the Boltzmann distribution of spin populations at finite temperature, naturally gives rise to non-classical behavior, even in near-equilibrium conditions~\cite{herzog2014boundary,schmidt2014using}. This fact has prompted ongoing efforts to characterize how signatures of nonclassical features manifest in thermally polarized NMR systems~\cite{hovav2013theoretical,khashami2023fundamentals}. In our earlier work~\cite{khashami2025quantum}, we explored quantum entanglement in thermally polarized, scalar-coupled two-spin-${1}/{2}$ systems using the concurrence measure. There, we examined the dependence of the entanglement measure on temperature and magnetic field strength, showing a threshold temperature above which all entanglement vanishes. While the analysis of concurrence provided a comprehensive understanding of quantum entanglement in the system, our primary focus was on quantifying entanglement in both homogeneous and heterogeneous systems. However, analyses beyond entanglement were not addressed.

The findings of Ref. ~\cite{khashami2025quantum} have motivated a broader investigation beyond entanglement, extending the framework to encompass other hallmarks of quantum behavior. To this end, the present study enhances this direction by shifting the emphasis toward quantum mixedness and quantum coherence to provide a more comprehensive analysis of quantumness of the system. Specifically, we examine the quantum coherence of the system through the relative entropy of the coherence \cite{}. Notably, quantum coherence has attracted considerable attention in the last decade due to the realization of quantum coherence as a resource \cite{baumgratz2014quantifying,streltsov2017colloquium}. Thus, we examine how quantum coherence and mixedness \cite{nielsen2010quantum} evolve with temperature, magnetic field strength, and interaction asymmetry. This extended analysis provides a comprehensive characterization of thermal quantum states and quantum features in them \cite{nielsen2010quantum,horodecki2009quantum,maleki2021quantum,jones2001nmr,oliveira2011nmr}. 
Our analysis connects NMR spectral features under scalar coupling~\cite{donovan2014heteronuclear,vuichoud2015measuring,appelt2010paths} to quantum features~\cite{jones2011quantum,cory2000nmr}. 

Furthermore, it is quite well-known that entanglement measures such as concurrence and negativity \cite{plenio2005logarithmic,verstraete2001comparison} are not associated with a quantum observable, making the experimental quantification of the entanglement 
a challenging issue in experiment. It was shown that entanglement witnesses can be used to detect entanglements that correspond to Hermitian operators \cite{chruscinski2014entanglement,eisert2007quantitative,guhne2009entanglement,brandao2005quantifying,maleki2018witnessing, terhal2000bell,oliveira2005complexity}. 
Therefore, one can detect entanglement using a Hermitian operator, corresponding to a quantum observable,  through entanglement witnesses \cite{chruscinski2014entanglement,eisert2007quantitative,guhne2009entanglement,brandao2005quantifying,maleki2018witnessing, terhal2000bell,oliveira2005complexity}, facilitating the experimental analysis of the entanglement in a quantum system. Having this in mind, we extend the entanglement analysis of NMR systems in Ref. ~\cite{khashami2025quantum}  to the analysis of the entanglement witnesses within this system. We show that the expected value of the energy of the system can be used to detect entanglement in the system, demonstrating that the entanglement witness can facilitate the experimental analysis of the entanglement in the NMR systems.

Therefore, this study complements earlier entanglement analysis by extending the analysis to quantum coherence and mixedness and analyzing the detection of quantum entanglement using entanglement witnesses. These studies collectively offer a unified perspective on quantum features and nonclassical behaviors in NMR systems, shedding light on both theoretical insight and experimental accessibility.

\section{Methods}

\subsection{Hamiltonian Description of a Thermally Polarized Two-Spin-1/2 NMR System}

\begin{figure}[h]
\centering  \includegraphics[width=\linewidth]{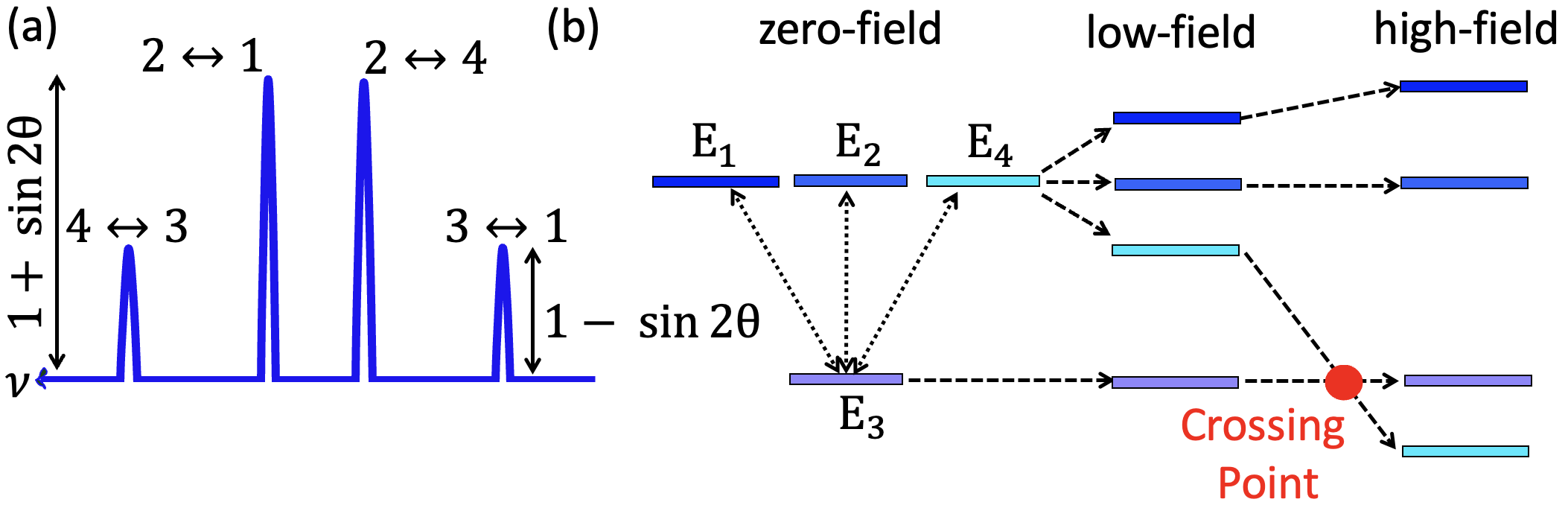}
\caption{(a) NMR spectrum for two-spin-$1/2$. (b) Field-dependent behavior of a homonuclear system,  magnetic field strength changes from zero-field to low-field and finally reaches high-field.  A level crossing occurs between $\mathrm{E}_3$ and $\mathrm{E}_4$, indicating a potential quantum transition point.}
\label{ground_state_nmr}
\end{figure}

We consider a pair of spin-${1}/{2}$ nuclei coupled via scalar interaction and subjected to a static external magnetic field ${B}_0$. The system's Hamiltonian is given by \cite{khashami2023fundamentals, mamone2020singlet, donovan2014heteronuclear,rudowicz2001spin}
\begin{equation}
\mathcal{H} = \mathcal{H}_{\text{Z}} + \mathcal{H}_{{\boldsymbol{J}}}, \label{Hamiltonian}
\end{equation}
where the Zeeman contribution is $\mathcal{H}_{\text{Z}} = (\omega_1 \sigma_{1z} + \omega_2 \sigma_{2z})/2$, with $\sigma_{iz}$ denoting the Pauli $z$ matrices and $\omega_i$ the Larmor frequency of spin $i$. The scalar coupling term is defined as $\mathcal{H}_{{\boldsymbol{J}}} = {{\boldsymbol{J}}}\,{\text{I}}_1 \cdot {\text{I}}_2$, where ${\text{I}}_i = {\sigma}_i/2$ is the spin angular momentum operator and ${\boldsymbol{J}}$ is the scalar coupling constant \cite{vuichoud2015measuring, ivanov2022chemically}.
To analyze the system, we diagonalize the total Hamiltonian based on the standard product $\{ |\alpha\alpha\rangle,\,|\alpha\beta\rangle,\,|\beta\alpha\rangle,\,|\beta\beta\rangle \}$. 
The resulting eigenstates take the form
\begin{equation} 
\begin{aligned}\label{eigenstates}
|\phi_1\rangle &= |\alpha\alpha\rangle, \quad\quad
|\phi_2\rangle = \cos\theta\,|\alpha\beta\rangle + \sin\theta\,|\beta\alpha\rangle, \\
|\phi_3\rangle &= -\sin\theta\,|\alpha\beta\rangle + \cos\theta\,|\beta\alpha\rangle, \quad\quad
|\phi_4\rangle = |\beta\beta\rangle,
\end{aligned}    
\end{equation}
where the mixing angle $\theta$ reflects the interplay between spin asymmetry and coupling strength, where $\sin(2\theta) = {{\boldsymbol{J}}}/{D}$ with $D = \sqrt{\omega_\delta^2 + {\boldsymbol{J}}^2}$, and $\omega_\delta = \omega_1 - \omega_2$ is the Larmor frequency difference. The corresponding energy eigenvalues in this basis are \cite{eykyn2021extended,khashami2023fundamentals}
\begin{equation}
\begin{aligned}
E_1 &= \frac{1}{2}(\omega_{\Sigma} + \frac{1}{2}{\boldsymbol{J}}), \\   
E_2 &= \frac{1}{2}(D - \frac{1}{2}{\boldsymbol{J}}), \\
E_3 &= -\frac{1}{2}(D + \frac{1}{2}{\boldsymbol{J}}), \\
E_4 &= \frac{1}{2}(-\omega_{\Sigma} + \frac{1}{2}{\boldsymbol{J}}),
\end{aligned} \label{EnergyLevel}  
\end{equation}
where $\omega_\Sigma = \omega_1 + \omega_2$. The outer states $|\alpha\alpha\rangle$ and $|\beta\beta\rangle$ are product states with parallel spins and are energetically shifted upward by $+{\boldsymbol{J}}/4$ relative to the center. In contrast, the inner states $|\phi_2\rangle$ and $|\phi_3\rangle$ are entangled superpositions of $|\alpha\beta\rangle$ and $|\beta\alpha\rangle$ that share a total $z$-projection of zero. These form a mixed doublet centered around $0$ and split by an effective energy gap $D$, with additional shifts of $\pm {\boldsymbol{J}}/4$. This level structure gives rise to the characteristic four-peak pattern observed in the NMR spectrum [see Fig. \ref{ground_state_nmr}(a)], with frequency increasing from right to left. Transitions are governed by single spin flips, allowing only transitions between adjacent eigenstates. As a result, the spectrum features four allowed lines such as $4 \leftrightarrow 3$, which denotes a transition between states with energies $E_4$ and $E_3$, corresponding to the dashed transitions between levels (see Fig. \ref{ground_state_nmr}(b)). Moreover, the signal intensities follow the characteristic roofing effect, which is modulated by the mixing angle $\theta$, resulting in amplitudes proportional to $1 \pm \sin 2\theta$ \cite{vuichoud2015measuring,khashami2023fundamentals}.

\subsection{Thermal State Representation of the Two-Spin-$1/2$ NMR System}

For a coupled system of two-spin-${1}/{2}$ at thermal equilibrium, the density matrix in its diagonal representation can be expressed as $\rho = \sum_{i=1}^n p_i \left|\phi_i\right\rangle \left\langle\phi_i\right|$, 
where the populations $ p_i $ satisfy $ p_i \geq 0 $ and $ \sum_i p_i = 1 $, ensuring the physical requirement $ \mathrm{Tr}\rho = 1 $. In thermal equilibrium, the state occupation probabilities follow the Boltzmann distribution $p_i = {e^{-\beta E_i}}/{Z}$, 
where $ \beta = 1/k_B T $ is the inverse thermal energy, with $ k_B $ the Boltzmann constant and $ T $ the absolute temperature. Moreover, the partition function is $Z = \sum_{i=1}^4 e^{-\beta E_i}$, 
which can be written in closed form as \cite{khashami2025quantum}
\begin{equation}
Z = 2  e^{ \frac{\beta {\boldsymbol{J}}}{4}}(e^{-\frac{\beta {\boldsymbol{J}}}{2}} \cosh (\frac{\beta \omega_{\Sigma}}{2}) + \cosh (\frac{\beta D}{2})). \label{partitionFunction}  
\end{equation}
The elements of the density matrix in the eigenbasis $\{|\phi_i\rangle\}$ are thus given by
\begin{align}
\rho_{11} &= \frac{1}{Z} e^{-\beta E_1}, \quad
\rho_{22} = \frac{1}{Z} (e^{-\beta E_2} \cos^2\theta + e^{-\beta E_3} \sin^2\theta), \nonumber\\
\rho_{33} &= \frac{1}{Z} (e^{-\beta E_2} \sin^2\theta + e^{-\beta E_3} \cos^2\theta), \quad
\rho_{44} = \frac{1}{Z} e^{-\beta E_4}, \nonumber \\
&\hspace{1em}\rho_{23} = \rho_{32} = \frac{1}{Z} (e^{-\beta E_2} - e^{-\beta E_3}) \sin\theta \cos\theta. 
\label{densityMatrixElement}
\end{align}
This construction ensures $ \sum_{i=1}^4 \rho_{ii} = 1 $,  $\rho_{22} \rho_{33} \geq\left|\rho_{23}\right|^2$ and $\rho_{11} \rho_{44} \geq\left|\rho_{14}\right|^2$.  The off-diagonal element $ \rho_{23} = \rho_{32} $, as given in Eq.~\eqref{densityMatrixElement}, where, the presence of non-zero off-diagonal terms as $ \rho_{23} $ reflects quantum coherence in the system states. This coherence terms can give rise to quantum entanglement in some settings. It turns out that the thermal state is entangled if and only if either \cite{zhang2024local,sanpera1998local,peres1996separability} 
\begin{equation}
  \rho_{11} \rho_{44} \leq\left|\rho_{23}\right|^2  \quad \text{or}  \quad \rho_{22} \rho_{33} \leq\left|\rho_{14}\right|^2.
\end{equation}
Otherwise, the state is separable  \cite{horodecki2001separability}.

Furthermore, the eigenvalues of the thermal state $\rho$ are given by~\cite{zhang2024local}
\begin{align}
\lambda_1 &= \frac{1}{2} [ (\rho_{11} + \rho_{44}) + \sqrt{(\rho_{11} - \rho_{44})^2} ] = \rho_{44}, \nonumber \\
\lambda_2 &= \frac{1}{2} [ (\rho_{11} + \rho_{44}) - \sqrt{(\rho_{11} - \rho_{44})^2}] = \rho_{11}, \nonumber \\
\lambda_3 &= \frac{1}{2} [ (\rho_{22} + \rho_{33}) + \sqrt{(\rho_{22} - \rho_{33})^2 + 4|\rho_{23}|^2}], \nonumber \\
\lambda_4 &= \frac{1}{2} [ (\rho_{22} + \rho_{33}) - \sqrt{(\rho_{22} - \rho_{33})^2 + 4|\rho_{23}|^2}].
\end{align}
The structure of $\rho$ determines the form of its eigenvalues. Since the energy eigenstates $|\phi_1\rangle$ and $|\phi_4\rangle$ are not coupled to any other eigenstate, the populations $\rho_{11}$ and $\rho_{44}$ directly yield two of the eigenvalues, namely $\lambda_1 = \rho_{44}$ and $\lambda_2 = \rho_{11}$. The remaining two eigenvalues arise from the coupled $\{|\phi_2\rangle, |\phi_3\rangle\}$, where the off-diagonal coherence $\rho_{23}$ introduces mixing. Diagonalizing the $2\times2$ submatrix formed by $\rho_{22}$, $\rho_{33}$, and $\rho_{23}$ leads to the eigenvalues $\lambda_3$ and $\lambda_4$. The square-root term reflects both the population imbalance, $|\rho_{22} - \rho_{33}|$,  and quantum coherence between $|\phi_2\rangle$ and $|\phi_3\rangle$, $|\rho_{23}|$. When $\rho_{23} = 0$, this reduces to the classical population splitting, when $\rho_{23}$ is large (e.g., for $\theta = \pi/4$), the eigenvalue gap becomes dominated by coherence. 
Given the energy ordering $ E_1 \geq E_4 $, as shown in Fig. \ref{ground_state_nmr}, it follows that $ \rho_{11} \leq \rho_{44} $, and hence $ \lambda_1 \geq \lambda_2 $.

\subsection{Field-Dependent Level Dynamics and Crossing Behavior}

To investigate how the magnetic field governs spin populations, spectral behavior, and quantum correlations, we examine the energy level evolution of a two-spin-$1/2$ NMR system as the field increases from zero to high values \cite{khashami2025quantum,carravetta2004beyond,miesel2006coherence,kimmich2004field,kiryutin2019transport}. At zero magnetic field, the Zeeman interaction vanishes, resulting in a degenerate triplet manifold ($E_1$, $E_2$, $E_4$) and a non-degenerate singlet ground state $E_3$, as illustrated in Fig.~\ref{ground_state_nmr}(b). This configuration thermally favors the population of the entangled singlet state at low temperature, yielding maximal thermal entanglement (concurrence $C = 1$), as reported in Ref.~\cite{khashami2025quantum}. As the magnetic field increases into the low-field regime, Zeeman splitting lifts the triplet degeneracy, causing the energy levels to separate. Notably, $E_3$ remains nearly stationary while $E_4$ decreases, leading to a level crossing between these two states.
The point of intersection, indicated as the "Crossing Point" in Fig.~\ref{ground_state_nmr}(b), corresponds to a quantum critical point in which the ground state changes its character from the entangled configuration $|\phi_3\rangle$ to the separable state $|\phi_4\rangle$ \cite{khashami2025quantum}. The crossing of the energy levels $E_3$ and $E_4$ which lead to a quantum critical point, occurs when the condition \cite{khashami2025quantum}
\begin{equation}
{\boldsymbol{J}} = \frac{\omega^2_{\Sigma}-\omega^2_{\delta}}{2\omega_{\Sigma}},
\label{crossingPoint}
\end{equation}
is satisfied. Equation~(\ref{crossingPoint}) serves as a unified expression valid for both homonuclear systems, where $\omega_1=\omega_2=\omega$, and heteronuclear systems, where $\omega_1 \neq \omega_2$ \cite{arnesen2001natural}. This general criterion establishes the boundary between entangled and separable ground states. Beyond this crossing, increasing the external field drives the system into a weak-coupling regime, with the level diagram and corresponding NMR spectral lines clearly indicating the suppression of entanglement.

\section{Results}

\subsection{Relative Entropy of Coherence For the Two-Spin-$1/2$ NMR System}

Quantum coherence \cite{baumgratz2014quantifying,streltsov2017colloquium} is a fundamental signature of quantum superposition and plays a critical role in quantum thermodynamics \cite{francica2020quantum,narasimhachar2015low,korzekwa2016extraction}, information processing  \cite{pan2017complementarity}, entanglement theory \cite{korzekwa2016extraction}, and quantum metrology \cite{maleki2021quantum,maleki2021quantum-2}. 
An important measure of quantum coherence is the relative entropy of coherence, which quantifies coherence by comparing the quantum state's entropy to that of its diagonal counterpart (which contains only classical probabilistic uncertainty). It is formally defined as \cite{baumgratz2014quantifying,streltsov2017colloquium}
\begin{equation}
\mathcal{R}(\rho) = S(\rho_d) - S(\rho),
\end{equation}
where $S(\rho) = -\mathrm{Tr}(\rho \log_2 \rho)$ denotes the von Neumann entropy of the density matrix $\rho$, and $\rho_d$ is the diagonal part of $\rho$ in the chosen reference basis.
For a two-qubit system, $\rho$ is a $4 \times 4$ density matrix, and $S(\rho)$ can be computed from its eigenvalues $\{\lambda_i\}_{i=1}^4$ as
\begin{equation}
S(\rho) = -\sum_{i=1}^{4} \lambda_i \log_2 \lambda_i.
\end{equation}
The diagonal entropy $S(\rho_d)$ is computed directly from the diagonal elements of $\rho$, is given by
\begin{equation}
S(\rho_d) = -\sum_{i=1}^{4} \rho_{ii} \log_2 \rho_{ii},
\end{equation}
where $\rho_{ii}$ denotes the population of the $i$-th basis state in the reference basis. Combining both expressions, the relative entropy of coherence becomes \cite{baumgratz2014quantifying,streltsov2017colloquium}
\begin{equation}
\mathcal{R}(\rho) = -\sum_{i=1}^{4} \rho_{ii} \log_2 \rho_{ii} + \sum_{i=1}^{4} \lambda_i \log_2 \lambda_i.
\end{equation}
This quantity is always non-negative ($\mathcal{R}(\rho)\ge 0$) and vanishes if and only if the state $\rho$ is diagonal, i.e., completely incoherent in the chosen basis. The first sum measures the classical uncertainty of the population distribution, the second contains the true quantum uncertainty of the full state. Their difference isolates how much entropy is purely due to quantum coherence.

\begin{figure}[h]
\centering
\includegraphics[width=\linewidth]{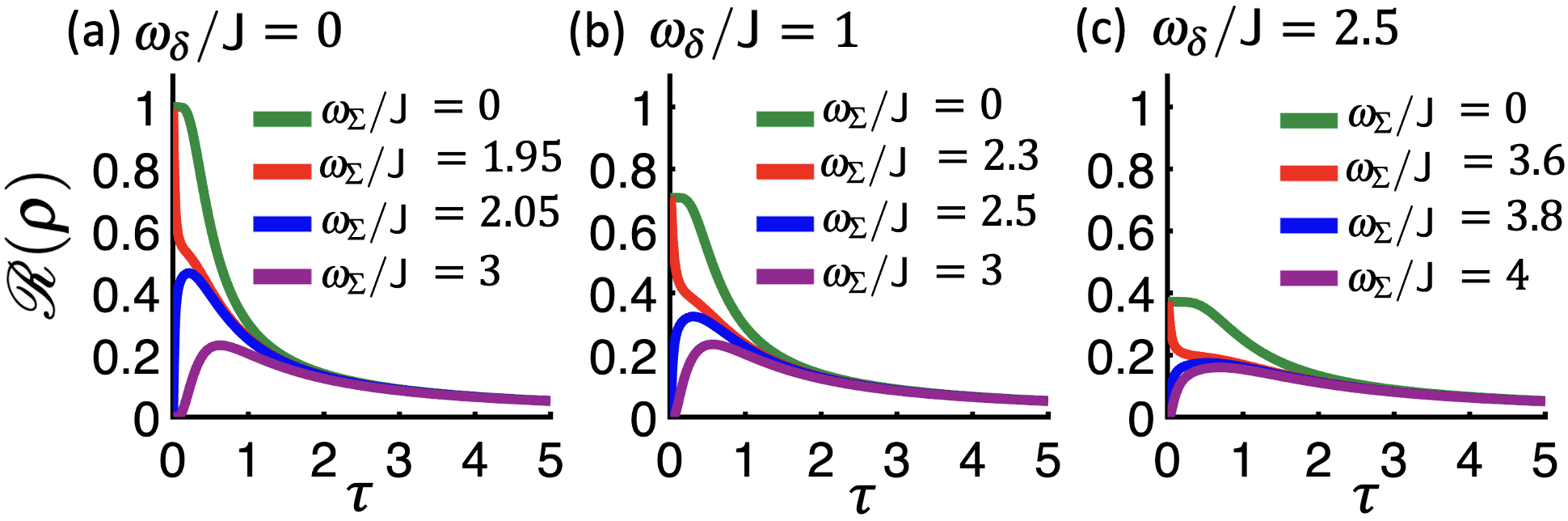}
  \caption{Relative entropy of coherence, $\mathcal{R}(\rho)$, versus the rescaled temperature for different values of the dimensionless magnetic field parameter $\omega_{\Sigma}/{\boldsymbol{J}}$. Panels (a), (b), and (c) correspond to $ \omega_{\delta}/{\boldsymbol{J}} = 0$ (a homonuclear system), $ \omega_{\delta}/{\boldsymbol{J}} = 1$ (a heteronuclear system), and $ \omega_{\delta}/{\boldsymbol{J}} = 2.5$ (a heteronuclear system), respectively. The re-scaled temperature parameter is $\tau = k_B T /{\boldsymbol{J}}$.}
\label{fig_RelativeEntropy}
\end{figure}

\begin{figure}[h]
\centering
  \includegraphics[width=\linewidth]{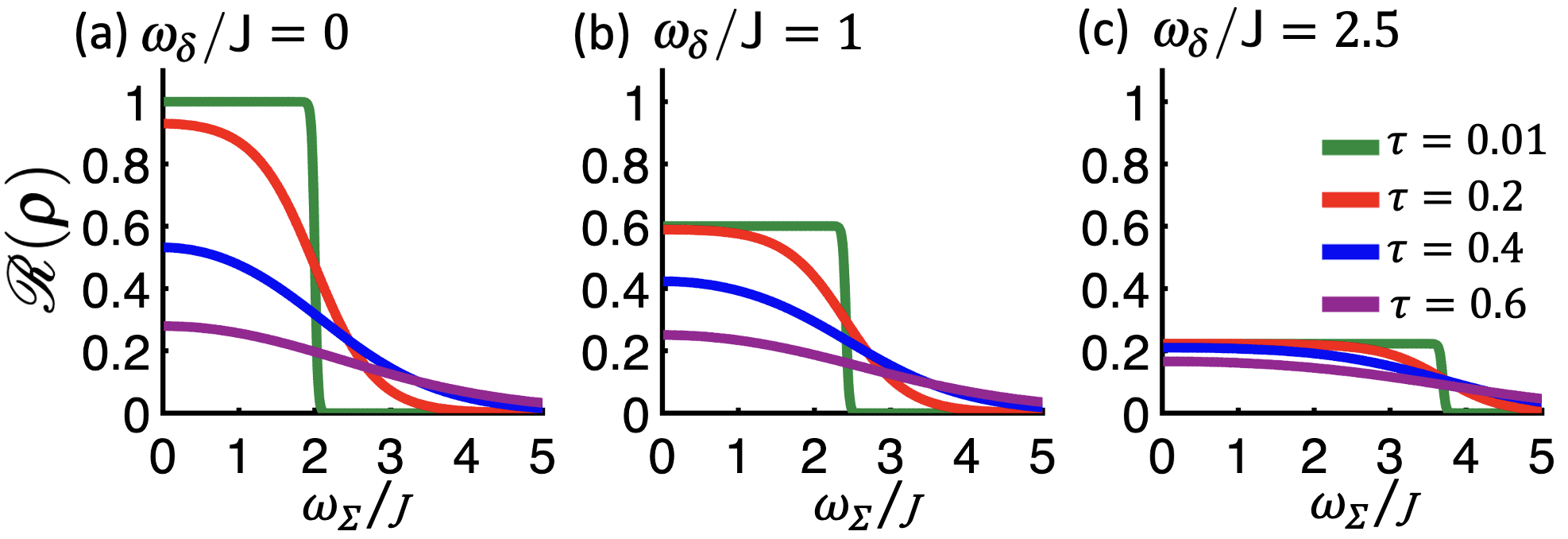}
  \caption{Temperature dependence of  $\mathcal{R}(\rho)$ for different values of the dimensionless magnetic field parameter $\omega_{\Sigma}/{\boldsymbol{J}}$. Panels (a), (b), and (c) correspond to $ \omega_{\delta}/{\boldsymbol{J}} = 0$ (a homonuclear system), $ \omega_{\delta}/{\boldsymbol{J}} = 1$ (a heteronuclear system), and $ \omega_{\delta}/{\boldsymbol{J}} = 2.5$ (a heteronuclear system), respectively. 
  }
	\label{rho_entropy_tem}
\end{figure}

The behavior of thermal quantum coherence under varying temperature and magnetic field conditions is illustrated in Figs.~\ref{fig_RelativeEntropy} and~\ref{rho_entropy_tem}. In Fig.~\ref{fig_RelativeEntropy}, $\mathcal{R}(\rho)$ is plotted as a function of the rescaled temperature $\tau = k_B T/{\boldsymbol{J}}$ for different values of $\omega_\Sigma/{\boldsymbol{J}}$, with each panel corresponding to a different value of  $\omega_\delta/{\boldsymbol{J}}$. Panel (a) represents the homonuclear case ($\omega_\delta/{\boldsymbol{J}} = 0$), and panels (b) and (c) show increasingly heteronuclear regimes with $\omega_\delta/{\boldsymbol{J}} = 1$ and $2.5$, respectively. 
At low temperatures, $\mathcal{R}(\rho)$ remains high when $\omega_{\Sigma}/\boldsymbol{J}$ is small and decreases monotonically with increasing temperature (green and red curves), reflecting the thermal degradation of quantum coherence. As $\omega_{\Sigma}/\boldsymbol{J}$ increases (blue and purple curves), $\mathcal{R}(\rho)$ starts from a lower value at low temperature and increases to maximum, then continues to decay as the temperature increases. This trend indicates that stronger magnetic fields and greater spin asymmetry reduce the degree of quantum superposition in the thermal state, leading to a faster loss of coherence with temperature.

Moreover, the overall magnitude of $\mathcal{R}(\rho)$ decreases with increasing $\omega_\delta/\boldsymbol{J}$, highlighting the fact that spin asymmetry suppresses mixing between  $|{\alpha\beta}\rangle$ and $|{\beta\alpha}\rangle$ states, thereby reducing the off-diagonal structure that quantifies coherence. The non-analytical behavior of quantum coherence at zero temperature characterizes the quantum critical point. We should note that similar 
 non-analytical behavior was observed in terms of concurrence as the measure of entanglement \cite{khashami2025quantum}.

Figure~\ref{rho_entropy_tem} complements this analysis by exploring the behavior of $\mathcal{R}(\rho)$ as a function of $\omega_\Sigma/{\boldsymbol{J}}$ at fixed temperatures $\tau$. Again, panels (a), (b), and (c) correspond to increasing spin asymmetry. In the low-temperature regime ($\tau = 0.01$), $\mathcal{R}(\rho)$ begins near its maximum value and decreases sharply with increasing $\omega_\Sigma$, capturing how strong external fields suppress quantum coherence. As the temperature increases, the decrease becomes less abrupt and the curves flatten, consistent with thermal averaging. In particular, the magnitude of $\mathcal{R}(\rho)$ is lower in the heteronuclear cases, confirming that spin asymmetry and magnetic field gradients act cooperatively to suppress coherence. 

%\subsubsection{Comparison Between Relative Entropy of Coherence and Concurrence}

In our previous study on thermal two-spin-$1/2$ NMR systems \cite{khashami2025quantum}, we employed concurrence \cite{wootters1998entanglement,hill1997entanglement,maleki2021naturalennett}  as a measure of entanglement to identify sharp threshold behaviors and quantum critical points associated with spin-spin coupling and magnetic field strength. While concurrence successfully captured the onset and disappearance of entanglement, it proved limited to capture the quantum features where the quantum state remained separable but still exhibited nonclassical features. The relative entropy of coherence, $\mathcal{R}(\rho)$, which quantifies quantum superposition in a chosen basis, can be nonzero even for states that with zero entanglement. This broader nonclassicality makes coherence an ideal candidate for probing quantum behavior in regimes where concurrence fails to respond.
While concurrence exhibits a sharp cutoff dropping to zero at a specific threshold temperature $T_t$, as reported in \cite{khashami2025quantum}, a phenomenon known as sudden death \cite{yu2009sudden,cunha2007geometry}, coherence behaves differently. It does not vanish at any finite value of temperature. In other words, even when the concurrence vanishes at $T_t$, the entropy of coherence $\mathcal{R}(\rho)$ remains finite and displays a nonzero value (see Fig.~\ref{fig_RelativeEntropy}). These measures distinguish nonlocal entanglement from basis-dependent coherence, offer a comprehensive view of the quantum structure in the interacting two-spin-$1/2$ system.

\subsection{Mixedness For the Two-Spin-$1/2$ NMR System}

When an NMR system is prepared in one of the eigenstates of the Hamiltonian in Eq.~(\ref{eigenstates}), it is described by a pure quantum state. In contrast, in thermal equilibrium, the system occupies multiple energy levels with Boltzmann distributed probabilities, leading to a mixed state described by the density matrix $\rho$. In two-spin-$1/2$ systems, mixedness quantifies how much this thermal occupation, along with the influence of external magnetic fields, degrades the purity of the quantum state and introduces statistical uncertainty~\cite{khashami2023fundamentals,adesso2004extremal}. Understanding how the magnetic field and temperature influence this mixedness is essential to characterize the quantum nature of the system state.

The mixedness of a quantum system can be quantified using the purity $\mathrm{Tr} \, \rho^2$, which measures the degree of statistical uncertainty in the system. For a Hilbert space of dimension $d$, this quantity ranges from $1/d$ for a maximally mixed state to $1$ for a pure state~\cite{nielsen2010quantum}. A natural choice for quantifying mixedness is therefore $1 - \mathrm{Tr} \rho^2,$
which vanishes for pure states and increases monotonically as the system becomes more mixed, reaching the maximum value of $1 - 1/d$ for a maximally mixed state. Rescaling this quantity yields the degree of mixedness as \cite{nielsen2010quantum}
\begin{equation}
\mathcal{M} = \frac{d}{d - 1} (1 - \mathrm{Tr} \, \rho^2 ), \label{mixedness}
\end{equation}
so that $\mathcal{M} = 0$ for a pure state and $\mathcal{M} = 1$ for a maximally mixed state. This measure effectively characterizes the deviation from being a pure state and can capture the transition from a pure ground state at very low temperatures to a thermally disordered mixed state as temperature increases in quantum systems.
In our system, the Hilbert space has dimension $d = 4$, and the purity is given by $\mathrm{Tr} \, \rho^2 = \sum_i e^{-2\beta E_i} / Z^2$. Substituting into Eq.~(\ref{mixedness}), the mixedness can be quantified as
\begin{equation}
\mathcal{M} = \frac{4}{3} (1 - \frac{1}{Z^2} \sum_i e^{-2\beta E_i}). \label{eq:mixedness_basic}
\end{equation}
Using the partition function from Eq.~(\ref{partitionFunction}), $\mathcal{M}$ reduces to 
\begin{equation}
\mathcal{M} = \frac{4}{3} \left[ 1 - \frac{e^{-\beta {\boldsymbol{J}}} \cosh(\beta \omega_\Sigma) + \cosh(\beta D)}{2 ( e^{-\beta {\boldsymbol{J}} / 2} \cosh( \frac{\beta \omega_\Sigma}{2}) + \cosh( \frac{\beta D}{2}) )^2}\right]. \label{eq:mixedness_closed}
\end{equation}
This expression captures how the mixedness depends on $T$, ${\boldsymbol{J}}$, $\omega_\Sigma$, and $D$. The numerator contains terms corresponding to energy-weighted contributions from both $\omega_\Sigma$ and detuned spin states, while the denominator involves the square of a sum that encodes thermal populations of these states. As $T$ increases, the denominator increases more rapidly than the numerator, leading to a higher value of $\mathcal{M}$ indicative of a greater statistical mixture and reduced quantum purity in the system.

It is essential to note that mixedness and concurrence, as reported in \cite{khashami2025quantum}, represent distinct aspects of a quantum state. Mixedness quantifies the overall purity of the two-spin-$1/2$ system, indicating how far the global state deviates from being pure. In contrast, concurrence measures the quantum entanglement between the two spins, characterizing their pairwise quantum correlations. While mixedness is a global property of the density matrix, concurrence focuses specifically on inter-particle entanglement.

\begin{figure}[h]
\centering
  \includegraphics[width=\linewidth]{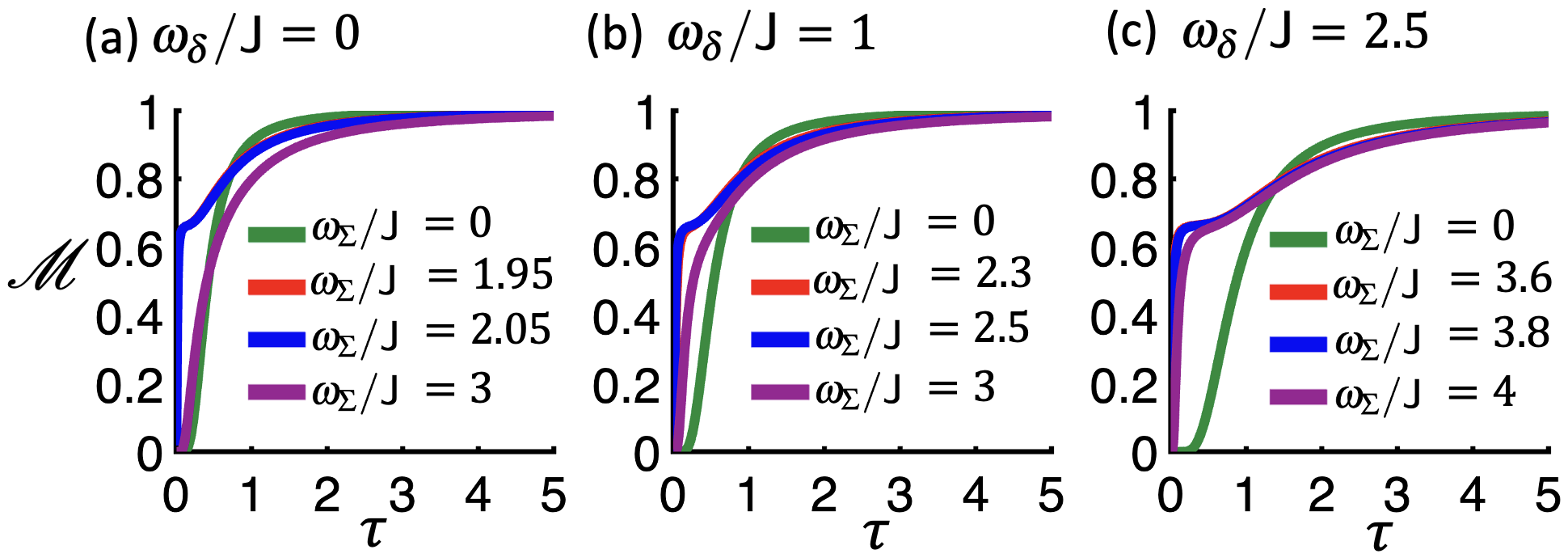}
  \caption{Mixedness as a function of the rescaled temperature for different values of the dimensionless magnetic field parameter $\omega_{\Sigma}/{\boldsymbol{J}}$. Panels (a), (b), and (c) correspond to $ \omega_{\delta}/{\boldsymbol{J}} = 0$ (a homonuclear system), $ \omega_{\delta}/{\boldsymbol{J}} = 1$ (a heteronuclear system), and $ \omega_{\delta}/{\boldsymbol{J}} = 2.5$ (a heteronuclear system), respectively. The re-scaled temperature parameter is $\tau = k_B T /{\boldsymbol{J}}$.}
  \label{Fig4_Mixedness}
\end{figure}

To illustrate this distinction, consider the pure eigenstates $|\phi_1\rangle$ and $|\phi_3\rangle$ defined in Eq.~(\ref{eigenstates}). Both are pure states, meaning their density matrices satisfy $\rho^2 = \rho$, and thus their mixedness vanishes, $\mathcal{M} = 1 - \mathrm{Tr}(\rho^2) = 0$. Despite having the same purity, their entanglement properties are fundamentally different. The state $|\phi_1\rangle$ is a triplet product state and fully separable, leading to zero concurrence ($C = 0$), indicating no quantum entanglement between the two spins. In contrast, the singlet state $|\phi_3\rangle$ is maximally entangled, yielding the  concurrence $C = 2 |\sin \theta \cos \theta|$. This contrast suggests that pure states can be either entangled or separable, and mixedness alone cannot determine entanglement. Mixedness quantifies statistical uncertainty or classical admixture in a quantum state, whereas concurrence shows nonlocal quantum correlations. Therefore, mixedness and concurrence serve complementary roles, one tracking the purity of the state, and the other identifying the presence or absence of entanglement.

Figure~\ref{Fig4_Mixedness} illustrates how $\mathcal{M}$ varies with the rescaled temperature $\tau$, showing an overall increase that reflects the loss of quantum purity due to thermal excitations and population mixing. The dependence of $\mathcal{M}$ on the ratio $\omega_\Sigma / {\boldsymbol{J}}$ highlights the interplay between internal spin interactions and external magnetic field control.
Specifically, when $\omega_\Sigma / {\boldsymbol{J}}$ is close to the quantum critical point, the mixedness rises sharply at low temperatures and then approaches the saturation value $\mathcal{M} \rightarrow 1$ gradually. In contrast, for $\omega_\Sigma / {\boldsymbol{J}}$ sufficiently smaller than the critical value, the growth in mixedness is slower at small $\tau$ but reaches saturation more quickly. When $\omega_\Sigma / {\boldsymbol{J}}$ is much larger than the critical value, the system tends to preserve purity over a wider temperature range, and the approach to saturation becomes significantly slower. Overall, the plot demonstrates the importance of tuning external field parameters to control thermal robustness and purity retention in spin-based quantum systems.

\begin{figure}[h]
\centering
  \includegraphics[width=\linewidth]{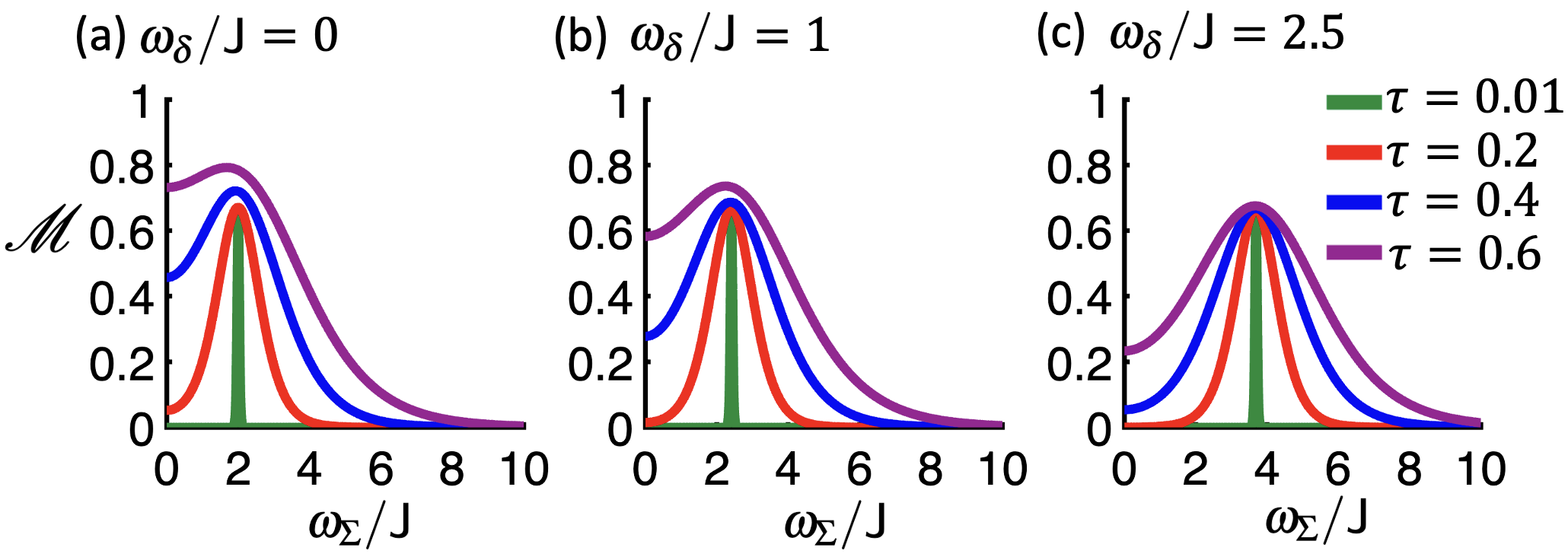}
  \caption{Mixedness as a function of the normalized frequency ratio $\omega_\Sigma / {\boldsymbol{J}}$ for varying values of the re-scaled temperature. Panels (a), (b), and (c) correspond to $ \omega_{\delta}/{\boldsymbol{J}} = 0 $ (a homonuclear system), $ \omega_{\delta}/{\boldsymbol{J}} = 1$ (a heteronuclear system), and $ \omega_{\delta}/{\boldsymbol{J}} = 2.5$ (a heteronuclear system), respectively, illustrating the influence of magnetic field strength and detuning on thermal mixedness. 
  }
  \label{Fig5_Mixedness}
\end{figure}

To further explore the role of external fields, we investigate how $\mathcal{M}$ varies with the rescaled frequency $\omega_\Sigma / {\boldsymbol{J}}$ at fixed values of the temperature. As shown in Fig.~\ref{Fig5_Mixedness}, $\mathcal{M}$ exhibits strong sensitivity to the external field configuration. Remarkably, the behavior of mixedness at zero temperature shows the location of the quantum critical point, marking the transition between distinct ground-state regimes. Furthermore, regardless of the value of $\omega_\Sigma / {\boldsymbol{J}}$, the mixedness of the system increases with temperature. This behavior is expected, as thermal excitation leads to increased population of higher-energy eigenstates, causing the overall state to become more statistically mixed. As the temperature rises, the populations of different eigenstates converge, reducing purity and increasing mixedness.
In the limit of negligible $\omega_\Sigma / {\boldsymbol{J}}$, the expression for mixedness simplifies to
\begin{equation}
\mathcal{M} \approx \frac{4}{3} \left[ 1 - \frac{e^{-\beta {\boldsymbol{J}}} + \cosh(\beta D)}{2 ( e^{-\beta {\boldsymbol{J}} / 2} + \cosh( \frac{\beta D}{2}) )^2} \right]. 
\end{equation}
In this setting, the mixedness when the system is homonuclear, i.e., $D = \boldsymbol{J}$, reduces to
\begin{equation}
\mathcal{M} \approx \frac{4}{3} \left[
1 - \frac{e^{2\beta {\boldsymbol{J}}}+3}{(e^{\beta {\boldsymbol{J}}}+2)^2}
\right].
\end{equation}
This result is consistent with the behavior shown in Fig.~\ref{Fig5_Mixedness}, where the mixedness is zero at zero temperature and increases with temperature. As thermal excitation becomes more significant, the system transitions from a pure to a mixed state, with $\mathcal{M}$ gradually approaching $1$ in the high-temperature limit.
In the regime where $\omega_\Sigma / {\boldsymbol{J}}$ is larger than the critical value, $\mathcal{M}$ decreases monotonically. In the high-field limit, the system tends to remain closer to a pure state over a wider temperature range. Moreover, within this region, the mixedness increases with increasing $D$, as the system becomes more asymmetric in terms of the spin Larmor frequencies. This asymmetry enhances thermal population mixing, contributing to a higher degree of mixedness.

To better understand the field-dependent behavior of thermal spin systems, we compare the mixedness results presented in this work (Fig.~\ref{Fig5_Mixedness}) with the concurrence profiles previously reported in Fig. 3~\cite{khashami2025quantum}. Concurrence measures quantum entanglement, while mixedness captures how thermal populations are distributed across the eigenstates. Despite probing different aspects of the quantum state, both parameters reflect the same underlying transition from coherent to incoherent behavior. In the homonuclear case ($\omega_\delta/{\boldsymbol{J}} = 0$), concurrence showed a sharp drop near $\omega_\Sigma/{\boldsymbol{J}} \approx 2$, which aligns with a pronounced peak in mixedness, marking the point of strongest spectral competition and loss of quantum correlations. As detuning increases ($\omega_\delta/{\boldsymbol{J}} = 1$ and $2.5$), both the concurrence threshold and the mixedness peak shift to higher field strengths. Entanglement becomes weaker overall, while the population spread broadens, indicating a more asymmetric Zeeman environment. Temperature has a similar effect on both, reducing concurrence and broadening the mixedness peak. These observations demonstrate that concurrence and mixedness provide complementary insights, with both capable of identifying the quantum critical point where the structure of the thermal state undergoes a fundamental change.

Interestingly, the thermal behavior in the mixedness plots (Fig.~\ref{Fig5_Mixedness}) shows the opposite trend to the concurrence plots from Fig.~3~\cite{khashami2025quantum}. As temperature increases, concurrence steadily decreases, showing the loss of quantum entanglement as the system moves toward a separable state. In contrast, mixedness increases with temperature, approaching $1$ at high temperatures as thermal energy spreads the population across more energy levels, making the system more statistically mixed.
Even though their trends can be opposite, both measures reflect the same process where rising temperature causes populations to spread among energy levels, reducing entanglement and increasing disorder. At low temperatures, concurrence is high and mixedness is low, meaning the system is nearly pure and entangled. As the temperature rises, concurrence drops to zero and mixedness reaches its maximum, showing a full transition from quantum order to thermal noise.
This inverse relationship highlights how the two measures complement each other. Concurrence reflects the strength of quantum correlations, whereas mixedness captures the loss of purity associated with decoherence. By combining both perspectives, we gain a more comprehensive understanding of how temperature drives the transition from quantum to classical behavior in two-spin-$1/2$ systems.

\subsection{NMR Signal Simulation based on Mixedness Analysis Near the Critical Point}

\begin{figure}[h]
\centering
  \includegraphics[width=\linewidth]{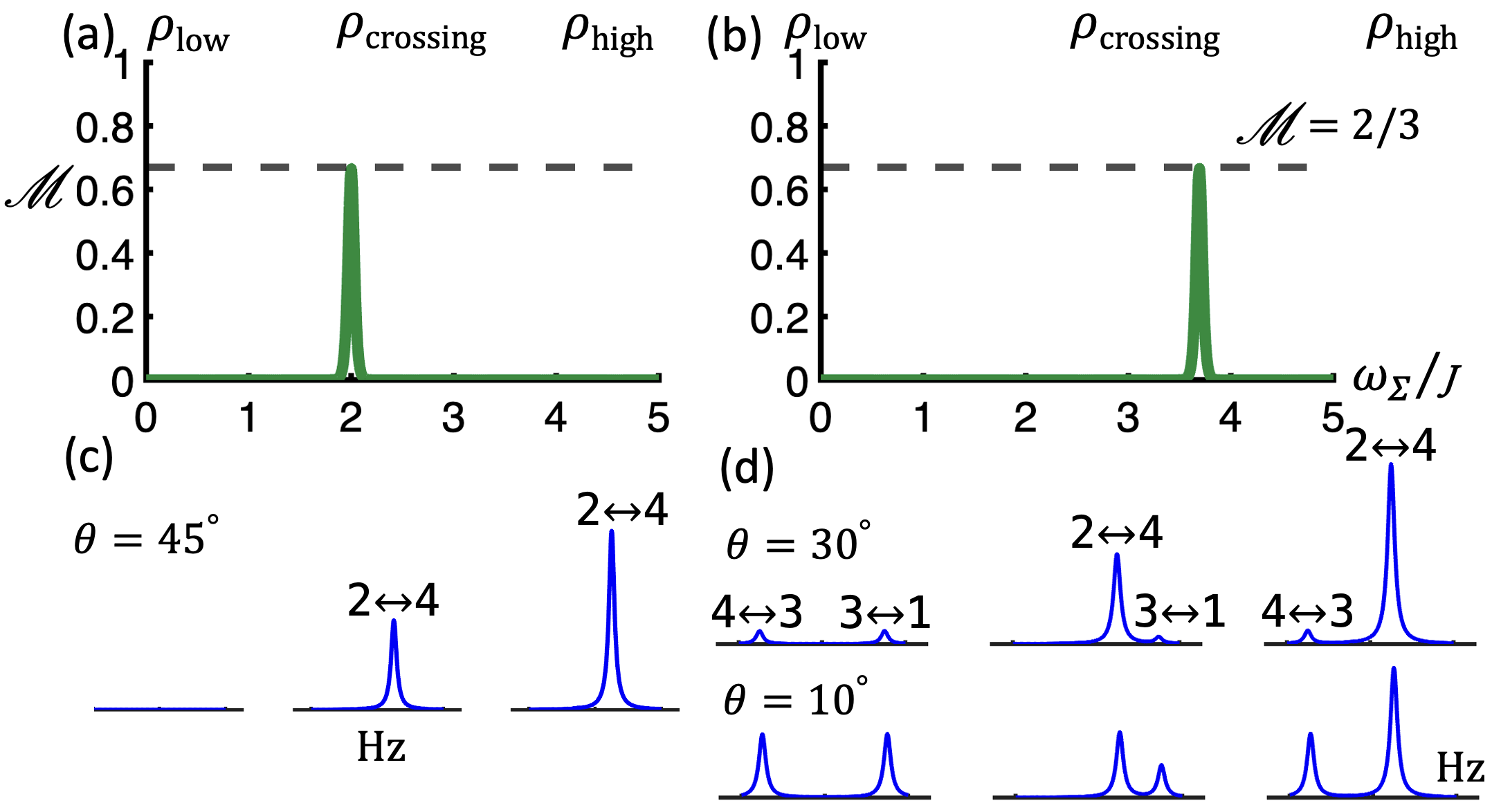}
\caption{Quantum signatures of ground‐state transitions. (a) Homonuclear case with $\omega_\delta/{\boldsymbol{J}}=0$, the mixedness $\mathcal{M}$ reaches its theoretical maximum $2/3$ at the population‐crossing point. (b) Heteronuclear case with $\omega_\delta/{\boldsymbol{J}}=2.5$, the mixedness peak shifts to larger $\omega_\Sigma/{\boldsymbol{J}}$ due to detuning. In both panels, the dashed line marks $\mathcal{M}=2/3$. (c–d) Simulated NMR spectra for $\rho_{\mathrm{low}}$, $\rho_{\mathrm{crossing}}$, and $\rho_{\mathrm{high}}$ with mixing angles $\theta=45^\circ$, $30^\circ$, and $10^\circ$ and flip angle $\varphi=5^\circ$ at $\tau = 0.01$. Peak labels (e.g., $2\leftrightarrow4$, $3\leftrightarrow1$) denote the allowed transitions.}
\label{ground_state}
\end{figure}

To investigate how ground-state degeneracy influences the mixedness of thermal states in NMR spin systems, we analyze the thermal density matrix in the near-zero temperature ($\tau = 0.01$). In this limit, the system does not necessarily settle into a unique pure state but may instead occupy a statistical mixture of degenerate ground states. To ensure consistency with our previous entanglement analysis based on concurrence \cite{khashami2025quantum}, we adopt the same computational framework to evaluate mixedness.
To capture spectral features linked to ground-state structure, we simulate transition amplitudes using the method described in Ref.~\cite{vuichoud2015measuring,khashami2025quantum}. These amplitudes are converted into Lorentzian lineshapes at the corresponding transition frequencies. For spectral simulations, we fix the flip angle at $\varphi = 5^\circ$ and set the mixing angle $\theta = 45^\circ$ for the homonuclear case, and $\theta = 30^\circ$ and $10^\circ$ for the heteronuclear case.  Our objective is to pinpoint the conditions under which the system transitions between pure and mixed states as a function of the magnetic field regime, and to characterize the associated quantum phase transition in a two-spin-$1/2$ system. Three representative scenarios emerge based on the order of the two lowest energy levels, for the low-field, the high-field, and the crossing point where $E_3 = E_4$, as shown in Fig. \ref{ground_state_nmr}.

In the low-field regime, the system fully relaxes into the eigenstate $ |\phi_3\rangle $, resulting in a pure non-degenerate regime characterized by zero mixedness ($ \mathcal{M} = 0 $), as shown in Fig. \ref{ground_state} (a)-(b). The corresponding thermal pure density matrix is given by
\begin{equation}
    \rho_{\mathrm{low}} = |\phi_3\rangle \langle \phi_3|.
\end{equation}
Under this condition, the only thermal transitions are $ 3 \leftrightarrow 1 $ and $ 4 \leftrightarrow 3 $. However, as shown in Fig.~\ref{ground_state}(c), when $ \theta = 45^\circ $ (homonuclear case), these transitions may be significantly suppressed or even symmetry-forbidden due to the structure of the eigenstates and selection rules imposed by the spin Hamiltonian \cite{mamone2020singlet}. As a result, the NMR spectrum for $ \rho_{\mathrm{low}} $ appears nearly flat, indicating negligible signal intensity. In contrast, Fig.~\ref{ground_state}(d) shows the heteronuclear case with smaller angle $ \theta = 30^\circ $ and $ \theta = 10^\circ $. At these lower angles, the mixing between spin states is weaker, and the eigenstates become more distinct. As a result, the transitions $ 3 \leftrightarrow 1 $ and $ 4 \leftrightarrow 3 $ now appear clearly in the spectrum with stronger intensity. This shows that in the low-field regime, detuning ($ \omega_\delta \neq 0 $) breaks the symmetry and reduces mixing between states, allowing more transitions to appear in the NMR spectrum.

At the crossing point, the two lowest energy states $ |\phi_3\rangle $ and $ |\phi_4\rangle $ are equally populated with $ p_3 = p_4 = 1/2 $, resulting in an intrinsically mixed ground-state with degeneracy. The corresponding thermal state is described by the density matrix $ \rho_{\mathrm{crossing}} $
\begin{equation}
  \rho_{\mathrm{crossing}}=\frac{1}{2}(|\phi_3\rangle \langle \phi_3| + |\phi_4\rangle \langle \phi_4|),  
\end{equation}
at the degeneracy point, $\mathrm{Tr}(\rho^2) = 1/2$, yielding a mixedness of 
\begin{equation}
 \mathcal{M} = \frac{2}{3} \approx 0.67.   
\end{equation}
This peak is prominently visible as a dashed line in Fig.~\ref{ground_state}(a) and (b), under the $\rho_{\mathrm{crossing}}$ condition. The corresponding NMR spectrum for the homonuclear case (Fig.~\ref{ground_state}(c), $ \theta = 45^\circ $) shows a dominant transition at $ 2 \leftrightarrow 4 $, and other transitions are suppressed due to symmetry and selection rules. In the heteronuclear case (Fig.~\ref{ground_state}(d)), at the crossing point with smaller mixing angles $ \theta = 30^\circ $ and $ \theta = 10^\circ $, the spectrum shows two main transitions, $ 2 \leftrightarrow 4 $ and $ 3 \leftrightarrow 1 $. This pattern reflects the reduced mixing between eigenstates at small angles, which modifies the transition amplitudes due to altered selection rules and weaker overlap between states. 

In the high-field limit, the system relaxes entirely into the lowest-energy state $ |\phi_4\rangle $, with population $ p_4 = 1 $ and all other levels unoccupied. The corresponding density matrix is
\begin{equation}
\rho_{\mathrm{high}} = |\phi_4\rangle \langle \phi_4|, 
\end{equation}
representing a pure state with no mixedness ($ \mathcal{M} = 0 $).
Under this condition, the only observable transitions are those connected to $ |\phi_4\rangle $, namely $ 4 \leftrightarrow 2 $ and $ 4 \leftrightarrow 3 $. 
In the homonuclear case (Fig.~\ref{ground_state}(c), $ \theta = 45^\circ $), the spectrum shows a single dominant peak at $ 2 \leftrightarrow 4 $, indicating that this transition carries most of the observable signal. In the heteronuclear case (Fig.~\ref{ground_state}(d)), the spectra for $ \theta = 30^\circ $ and $ \theta = 10^\circ $ show two distinct transitions, $ 2 \leftrightarrow 4 $ and $ 4 \leftrightarrow 3 $. As the mixing angle decreases, mixing between spin states weakens, causing the transition amplitudes to redistribute and the peaks to become less intense. 

Moreover, as the detuning ratio $ \omega_\delta/{\boldsymbol{J}} $ increases from zero (Fig.~\ref{ground_state}(a)) to $2.5$ (Fig.~\ref{ground_state}(b)), the location of the critical point shifts toward higher values of $ \omega_\Sigma/{\boldsymbol{J}} $ at the energy level crossing. This behavior reflects how detuning governs the onset of quantum criticality. The shift is described analytically by Eq.~(\ref{EnergyLevel}), where degeneracy arises when $ D/{\boldsymbol{J}} = \omega_\Sigma/{\boldsymbol{J}} - 1 $.
This transition is further confirmed in \cite{khashami2025quantum}, which shows that small changes in $ \omega_\Sigma/{\boldsymbol{J}} $ at low temperatures cause sudden drops in concurrence. These abrupt changes highlight the sensitivity of spin correlations to ground-state degeneracy and serve as practical indicators for identifying quantum phase transitions in NMR experiments.

\subsection{Experimental Protocol For Estimation of Mixedness in Two-Spin-$1/2$ NMR Systems}

Next, we extend analysis to propose an experimentally accessible protocol to quantify mixedness. This provides deeper insight into how thermal effects and ground-state degeneracy shape the structure of quantum states in realistic NMR systems.
Our method is based on three measurable NMR observables: the individual Zeeman polarizations, $\mathrm{P}_{1z}= \mathrm{Tr}(\sigma_{1z} \rho) $ and $\mathrm{P}_{2z}= \mathrm{Tr}(\sigma_{2z} \rho) $, and the longitudinal two-spin correlation $\mathrm{P}_{1z,2z}= \mathrm{Tr}(\sigma_{1z} \sigma_{2z}\rho) $ \cite{lu2016tomography,lvovsky2009continuous,gross2010quantum,xin2017quantum,vuichoud2015measuring}. 
To perform this protocol experimentally, three standard $1$D NMR experiments are required: two selective readouts to measure $\mathrm{P}_{1z}$ and $\mathrm{P}_{2z}$, and one additional sequence to extract $\mathrm{P}_{1z,2z}$. These measurements can be implemented using straightforward pulse sequences with appropriate pulse phases and refocusing delays, without requiring full quantum state tomography. The mixing angle $\theta$ is determined from the system Hamiltonian parameters, which are typically known from spectral calibration. The following observables are written as \cite{khashami2025quantum}
\begin{equation}
\begin{aligned}
\mathrm{P}_{1z} &= p_1 - p_4 + (p_2 - p_3) \cos 2\theta, \\
\mathrm{P}_{2z} &=  p_1 - p_4 + (p_3 - p_2) \cos 2\theta, \\
\mathrm{P}_{1z,2z} &=  p_1 + p_4 - (p_2 + p_3).
\end{aligned}
\label{eq_populations}
\end{equation}
Assuming $\cos 2\theta \neq 0$, the populations $\{p_i\}$ can be fully reconstructed as
\begin{equation}
\begin{aligned}
p_1 &= \frac{1}{4} (1 + \mathrm{P}_{1z} + \mathrm{P}_{2z} + \mathrm{P}_{1z,2z}), \\
p_2 &= \frac{1}{4} (1 - \mathrm{P}_{1z,2z} + \frac{\mathrm{P}_{1z} - \mathrm{P}_{2z}}{\cos 2\theta}), \\
p_3 &= \frac{1}{4} (1 - \mathrm{P}_{1z,2z} + \frac{\mathrm{P}_{2z} - \mathrm{P}_{1z}}{\cos 2\theta}), \\
p_4 &= \frac{1}{4} (1 - \mathrm{P}_{1z} - \mathrm{P}_{2z} + \mathrm{P}_{1z,2z}).
\end{aligned}
\label{eq_populations_corrected}
\end{equation}
Using the reconstructed populations, we evaluate the mixedness $\mathcal{M}$ of the thermal state via the linear entropy definition, $\mathcal{M} = 1 - \mathrm{Tr}(\rho^2)$. By substituting the expressions from Eq.~(\ref{eq_populations_corrected}), we derive a closed-form formula that expresses mixedness entirely in terms of experimentally accessible NMR observables as
\begin{equation}
\mathcal{M} = 1 - \frac{1}{6} \left[ 
2\,\mathrm{P}^2_{1z,2z} 
+ \left( \mathrm{P}_{1z} + \mathrm{P}_{2z} \right)^2 
+  \frac{\left( \mathrm{P}_{1z} - \mathrm{P}_{2z} \right)^2}{\cos^2 2\theta} 
\right]. \label{mixednessEq}
\end{equation}
These results demonstrate that, like concurrence, mixedness, often viewed as an abstract, non-observable property of quantum states, can be experimentally reconstructed using standard NMR measurements. Combing mixedness with the previously established entanglement framework enables a more comprehensive analysis of thermal state structure and provides a practical route for experimentally benchmarking quantum purity and correlations in coupled spin-$1/2$ systems.

\subsection{ Entanglement Detection via Witness Operators in Two-Spin-$1/2$ NMR Systems}

Entanglement witnesses \cite{chruscinski2014entanglement,eisert2007quantitative,guhne2009entanglement,brandao2005quantifying,maleki2018witnessing, terhal2000bell,oliveira2005complexity} provide a powerful means to bridge theoretical descriptions of quantum correlations with experimentally measurable observables. Unlike entanglement measures such as concurrence, which require full knowledge of the density matrix, witness operators can detect entanglement through a small set of expectation values, often accessible in standard experimental platforms. This makes them especially practical in NMR systems, where correlators like longitudinal two-spin operators are closely tied to entanglement. Here, Bell-type witness operators \cite{jeong2006greenberger} tailored to the singlet-like ground state offer a direct and experimentally feasible method for identifying entangled states. 

\begin{figure}[h]
\centering
\includegraphics[width=5.5cm]{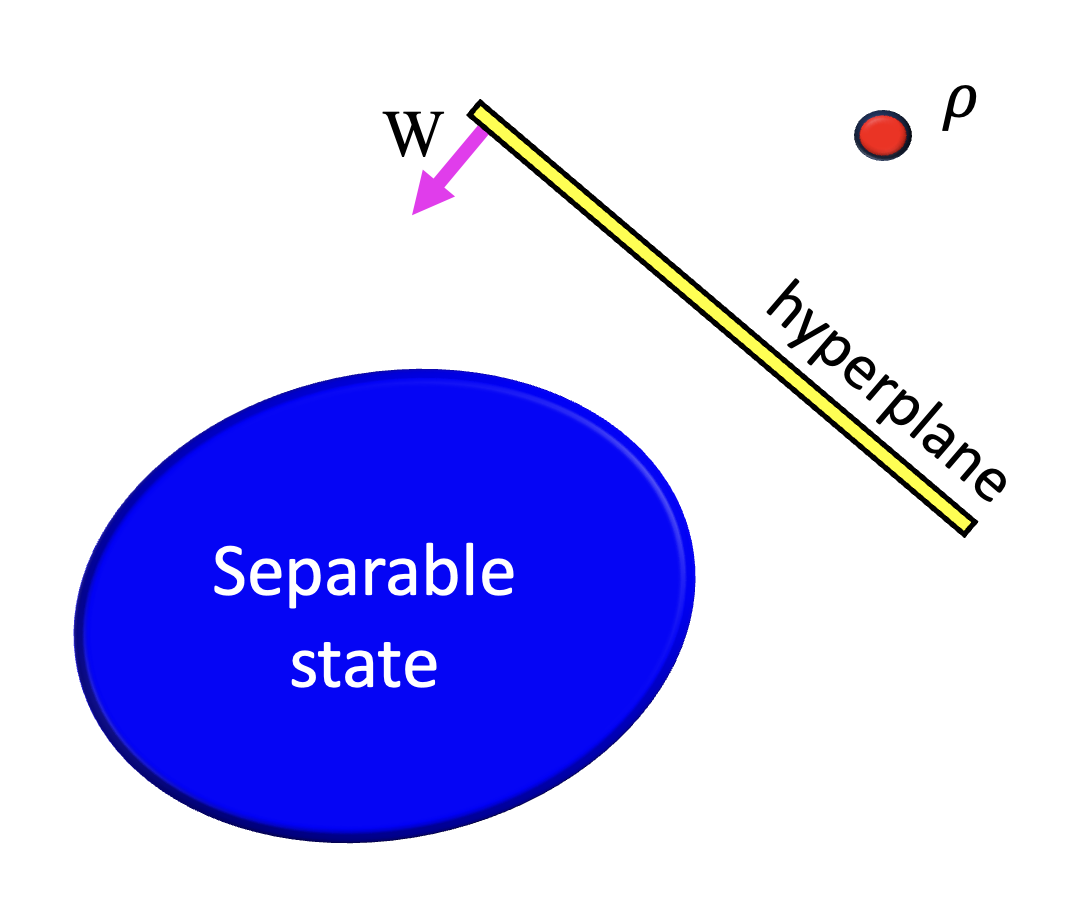}
\caption{Geometric representation of an entanglement witness. The enclosed region (blue area) represents the convex set of all separable states. The operator $W$ defines a hyperplane (yellow line) that separates this set from a subset of entangled states lying outside it. For all separable states $\rho$, the expectation value satisfies $\operatorname{Tr}(W\rho) \geq 0$, whereas for at least one entangled state $\rho$, the value becomes negative, $\operatorname{Tr}(W\rho) < 0$. }
\label{fig_entanglement_witness}
\end{figure}

The concept of an entanglement witness provides an elegant geometric picture for distinguishing entangled states from separable ones. In the state space of all density operators, the set of separable states forms a convex set, meaning that any statistical mixture of separable states remains separable. This region is depicted in Fig.~\ref{fig_entanglement_witness} as a closed, bounded domain in blue color.
An entanglement witness is a Hermitian operator $W$ acting on the bipartite Hilbert space $\mathbb{C}^2 \otimes \mathbb{C}^2$ that satisfies $\mathrm{Tr}(W \rho) \geq 0$ for all separable $\rho$, but $\mathrm{Tr}(W \rho) < 0$
for at least one entangled state. 
This condition implies that $W$ defines a hyperplane that separates at least one entangled state $\rho$ from the separable states. When this condition is met, the negativity of the expectation value serves as conclusive evidence that the tested state $\rho$ is entangled.  In the case of two-qubit systems, this geometric separation is always possible, for every entangled state, there exists some entanglement witness that detects it. Since entanglement witnesses are Hermitian operators, they are associated with physical observables and can be implemented in experiments using standard measurement techniques.

A canonical example of an entanglement witness is the one that targets the maximally entangled singlet state, denoted by $|\phi_3\rangle = (|\alpha \beta\rangle - |\beta \alpha\rangle)/{\sqrt{2}}$, when $\theta = \pi/4$ in  Eq. (\ref{eigenstates}). The associated witness for singlet eigenstate is written as  \cite{guhne2009entanglement}
\begin{equation}
W_{\text{singlet}} = \frac{1}{2} \mathbb{I} - |\phi_3\rangle \langle \phi_3|. \label{witphi3}
\end{equation}
To evaluate the expectation value of the witness operator in an arbitrary quantum state $\rho$, we compute
\begin{equation}
\langle W_{\text{singlet}} \rangle_\rho 
= \operatorname{Tr}(\rho\,W_{\text{singlet}}).
\end{equation}
Substituting the definition of $W_{\text{singlet}}$ from Eq.~\eqref{witphi3}, we obtain
\begin{equation}
\langle W_{\text{singlet}} \rangle_\rho = \operatorname{Tr}(\rho\,W_{\text{singlet}}) 
= \frac{1}{2} \operatorname{Tr}(\rho) - \operatorname{Tr}(\rho\,|\phi_3\rangle\langle\phi_3|),
\end{equation}
where the term $\operatorname{Tr}(\rho\,|\phi_3\rangle\langle\phi_3|) = \langle \phi_3 | \rho | \phi_3 \rangle$ is the fidelity \cite{liang2019quantum} between $\rho$ and the singlet state, $F( \rho, |\phi_3\rangle\langle \phi_3|) = \langle \phi_3 | \rho | \phi_3 \rangle$, so we have  
\begin{equation}
\langle W_{\text{singlet}} \rangle_\rho 
= \frac{1}{2} - F( \rho, |\phi_3\rangle\langle \phi_3|).
\label{eq_witness_fidelity}
\end{equation}
Thus, the inequality $\langle W_{\text{singlet}} \rangle_\rho < 0$ is equivalent to $F( \rho, |\phi_3\rangle\langle \phi_3|) > {1}/{2}$.

Thus, a quantum state $\rho$ is entangled if its fidelity with the singlet state exceeds ${1}/{2}$. This provides a simple, powerful, and sufficient condition for entanglement detection.
The witness operator $W_{\text{singlet}}$ can be decomposed into Pauli tensor products and evaluated through measurable spin correlators. This makes it a practical tool in quantum physics and NMR platforms where spin-spin correlations can be accessed through controlled rotations and longitudinal detection.
To connect this criterion to experimentally accessible quantities, the singlet projector can be expanded in the Pauli basis as \cite{guhne2009entanglement}
\begin{equation}
|\phi_3\rangle \langle \phi_3| = \frac{1}{4} \left( \mathbb{I} - \sigma_x \otimes \sigma_x - \sigma_y \otimes \sigma_y - \sigma_z \otimes \sigma_z \right).
\end{equation}
Substituting this form into the expression for the witness Eq. (\ref{witphi3}) yields
\begin{equation}
\langle W_{\text{singlet}} \rangle_\rho = \frac{1}{4} \left( 1 + C_{xx} + C_{yy} + C_{zz} \right), \label{witness}
\end{equation}
where $C_{\alpha\alpha} = \langle \sigma_{1\alpha} \sigma_{2\alpha} \rangle$ are the two-spin Pauli correlators along the $\alpha = x, y, z$ directions. The entanglement condition now reduces to the inequality, $\langle W_{\text{singlet}} \rangle_\rho < 0$, as
\begin{equation}
C_{xx} + C_{yy} + C_{zz} < -1, \label{inequality}
\end{equation}
which provides a practical test for entanglement in terms of measurable spin-spin correlations. If the sum of these three correlators drops below $-1$, the state is necessarily entangled. In a pulsed NMR experiment, these correlators can be measured through carefully designed rotation and readout sequences. The longitudinal two-spin correlator $C_{zz} = \langle \sigma_{1z} \sigma_{2z} \rangle$ is directly accessible from the standard readout, as it corresponds to the spin alignment along the $z$-axis. To measure $C_{xx}$ correlator, a ${\pi}/{2}$ pulse about the $y$-axis is applied to both spins  and the resulting signal is read out as $P_{1z,2z}$, which corresponds to the pre-rotation value of $C_{xx}$. Similarly, to measure $C_{yy}$ correlator, a ${\pi}/{2}$ pulse about the $-x$-axis is applied to both spins, and the same longitudinal readout $P_{1z,2z}$ yields the value of $C_{yy}$ before rotation. These three measured correlators are then substituted into the entanglement inequality (Eq. (\ref{inequality})) to test for nonclassical correlations.

The entanglement witness can be expressed as the energy quantities based on the Hamiltonian of the system. By removing the Zeeman term from the Hamiltonian in Eq. (\ref{Hamiltonian}) and normalizing by the coupling constant, we obtain an energy-based entanglement witness, expressed as      
\begin{equation}
W_{\text{singlet}} \equiv \frac{1}{4} \mathbb{I} + \frac{1}{\boldsymbol{J}}(\mathcal{H} - \mathcal{H}_{\text{Z}}).
\end{equation}
Then, for expected value of the witness we have
\begin{equation}
\langle W_{\text{singlet}} \rangle_\rho = \frac{1}{4} \mathbb{I} + \frac{1}{\boldsymbol{J}}(\langle \mathcal{H} \rangle - \langle \mathcal{H}_{\text{Z}} \rangle), \label{ham_wit}
\end{equation}
thus $\langle W_{\text{singlet}} \rangle_\rho < 0$, the $\rho$ is entangled. This witness is particularly useful in practice, as it depends only on pairwise spin correlations and excludes local field contributions that do not generate entanglement, making it well suited for entanglement certification in NMR experiments.

\begin{figure}[h]
    \centering
\includegraphics[width=\linewidth]{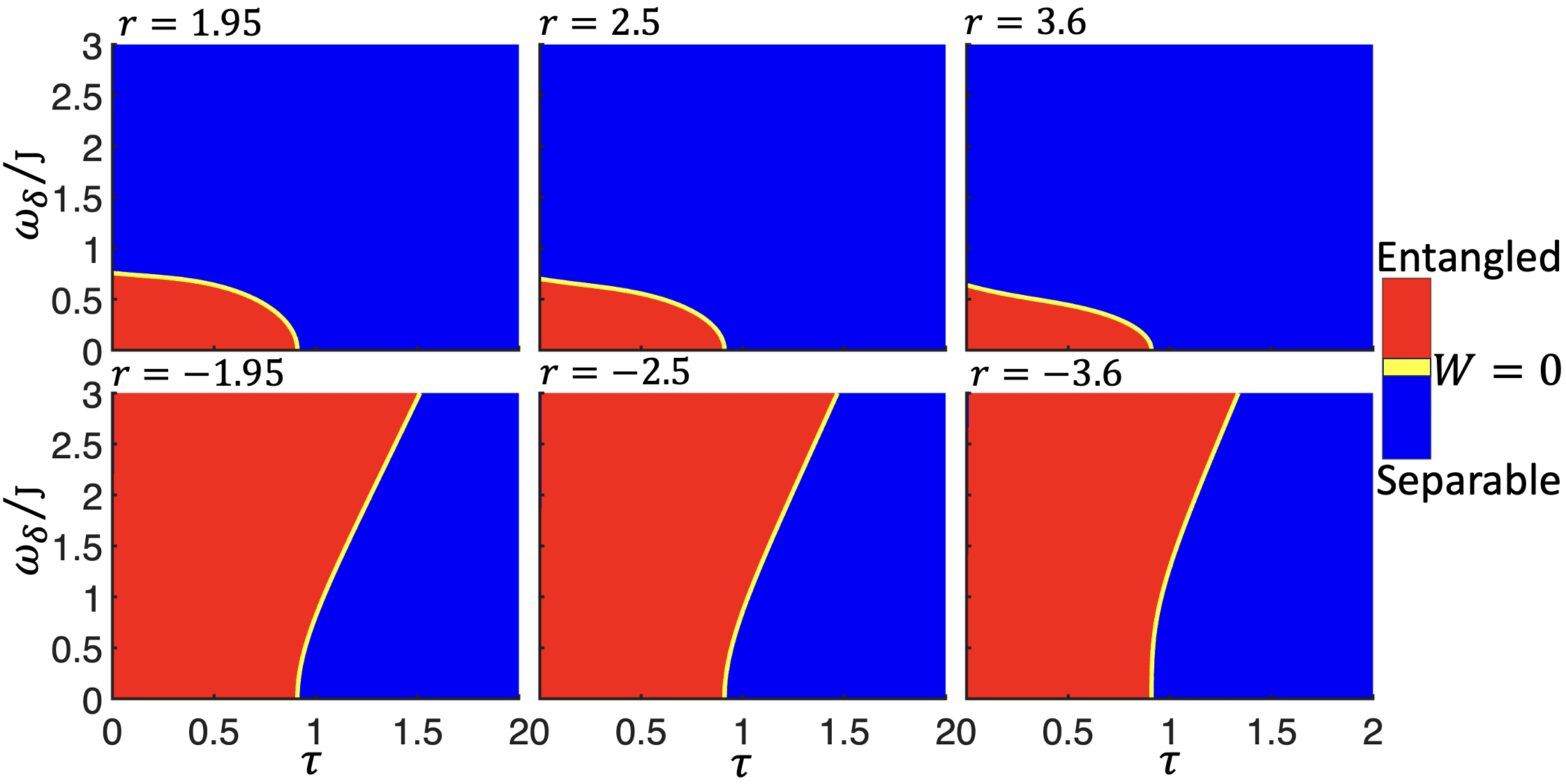}
\caption{Entanglement phase diagrams based on the singlet witness for various magnetic field ratios $r = B_1 / B_2$.  Red indicates entangled thermal states, blue indicates separable states, and the yellow contour marks the witness boundary $ \langle W_{\text{singlet}} \rangle_\rho = 0 $.}
\label{fig_spectra_wit}
\end{figure}

Figure~\ref{fig_spectra_wit} illustrates the entanglement–separability boundary obtained from the $ W_{\text{singlet}} $, plotted as a function of $ \tau = k_B T / {\boldsymbol{J}} $ and  $ \omega_\delta / {\boldsymbol{J}} $. The calculations are performed using Eq. (\ref{ham_wit}) for different magnetic field ratios $r = B_1 / B_2$ between the two spins, here $\omega = \gamma B$ and $\gamma$ is gyromagnetic ratio, which can be positive ($\omega$ parallel to $B$) or negative ($\omega$ anti-parallel to $B$). To explore how field asymmetry affects thermal entanglement, the simulation spans a range of $r$ values. For each case, $\langle W_{\text{singlet}} \rangle_\rho$ is computed over a grid of $(\tau, \omega_\delta / {\boldsymbol{J}})$ points. In each panel, the yellow contour marks the condition $\langle W_{\text{singlet}} \rangle_\rho = 0$, showing the entanglement–separability boundary, which can be computed analytically from Eq.~(\ref{witness}). 
The colormap is binary, consisting of two distinguishable colors, red for the entangled region ($\langle W_{\text{singlet}} \rangle_\rho < 0$) and blue for the separable region ($\langle W_{\text{singlet}} \rangle_\rho \ge 0$). Generally, in all planes, at low temperatures (close to $0$), quantum correlations are dominant, and a broad entangled phase emerges. As temperature increases, thermal fluctuations progressively suppress quantum coherence, reducing the extent of the entangled region. This leads to the eventual disappearance of entanglement, in agreement with the threshold temperature identified in Figure 2 \cite{khashami2025quantum}.

The plots in Fig.~\ref{fig_spectra_wit} show how thermal entanglement depends on the direction and strength of the local magnetic fields applied to the two spins. A key difference emerges when comparing positive and negative $r$, which highlights the importance of local field asymmetry. When \( r > 0 \), corresponding to spins with the same sign of gyromagnetic ratios, experiencing magnetic fields oriented in the same direction. This configuration breaks the entanglement of the system and favors separable product states, such as \( |\phi_1\rangle\) and \( |\phi_4\rangle\), where each spin aligns with its local field. These separable configurations dominate the thermal density matrix at moderate to high temperatures. Although spin--spin interactions still support entangled eigenstates like \( |\phi_2\rangle \) and \( |\phi_3\rangle \), which involve superpositions of \( |\alpha\beta\rangle \) and \( |\beta\alpha\rangle \), these states are suppressed as the external fields grow stronger. As a result, the entangled region in the witness plot becomes confined to a narrow window of low detuning and low temperature.

In contrast, for \( r < 0 \), two spins with opposite signs of gyromagnetic ratios, experiencing magnetic fields orienting spins in opposite directions. The field acting on one spin enhances alignment along \( +z \), while the other favors \( -z \). This competition enhances the stability of nonclassical superpositions, especially the singlet-like \( |\phi_3\rangle \) state. Importantly, the singlet state is invariant under equal and opposite field shifts, meaning that the antisymmetric field configuration does not destroy its entanglement. As a result, the entangled region in the \( r <0 \) plot expands significantly, occupying much of the detuning-temperature space, reflecting entanglement is more robust when local fields effects oppose one another.

\section{Discussions}

Extensive theoretical and experimental studies have explored entanglement in coupled spin-$1/2$ systems under thermal equilibrium and external magnetic fields \cite{wang2001entanglement,guo2003thermal,asoudeh2005thermal,vcenvcarikova2020unconventional,adamyan2020quantum}. In our recent work \cite{khashami2025quantum}, we established how temperature, field strength, anisotropy, and spin–spin interactions govern the degree of thermal entanglement in two-spin-$1/2$ NMR system. Other central quantum quantities of a system include quantum coherence and mixedness, which can shed light on the quantum nature of the systems beyond entanglement \cite{baumgratz2014quantifying,streltsov2017colloquium,francica2020quantum,narasimhachar2015low,korzekwa2016extraction}. Quantum coherence, a central feature of quantum mechanics and a key element in quantum resource theories, quantifies the degree to which a state exists as a superposition rather than a classical mixture \cite{pan2017complementarity,korzekwa2016extraction,maleki2021quantum,berrada2025quantum}. Mixedness, in contrast, characterizes the statistical uncertainty of a quantum state, often evaluated through measures such as linear entropy \cite{nielsen2010quantum}.  Therefore, it is quite important to analyze these quantities in coupled spin-$1/2$ NMR systems, and consider the interplay in systems where coherence, purity, and entanglement coexist under experimentally tunable thermal and magnetic conditions.
Building upon the previous entanglement-based framework \cite{khashami2025quantum}, this work introduces an experimentally accessible approach for quantifying mixedness, coherence entropy, and entanglement witness in two-spin systems using standard NMR polarization observables. This unified treatment allows us to explore the interplay between state purity, coherence, and entanglement within a consistent thermodynamic and magnetic framework. By expressing these quantifiers in terms of measurable NMR spectra, we demonstrate that quantities traditionally regarded as abstract theoretical constructs can be experimentally reconstructed in a chemically relevant and tractable manner.

\section{Conclusion}

This study extends the quantum information analysis of spin-$1/2$ NMR systems by emphasizing three key descriptors, mixedness, coherence entropy, and entanglement witness. Building on earlier work that examined entanglement via concurrence \cite{khashami2025quantum}. We present a unified framework that integrates entanglement and purity based quantifiers, thereby deepening the understanding of how thermal and magnetic parameters shape the quantum structure of coupled spins. The derived analytical relations show that mixedness and coherence display strong temperature dependence, with a non-analytic transition at zero temperature marking a quantum critical point. By introducing an entanglement witness directly expressible in terms of experimentally measurable NMR observables, we provide a practical route for detecting and quantifying entanglement in real spin systems. This approach demonstrates that key quantum information metrics can be reconstructed from standard polarization measurements without requiring full quantum state tomography. The resulting correspondence between theoretical quantifiers and measurable NMR spectra establishes a bridge between quantum information science and magnetic resonance spectroscopy. Overall, this framework offers both theoretical and experimental pathways for characterizing quantum correlations, purity, and decoherence in coupled-spin systems. It lays the groundwork for future investigations of thermal quantum effects, entanglement control, and quantum thermometry in NMR and related spectroscopic platforms, reinforcing the emerging synergy between quantum information theory and experimental magnetic resonance.

\section*{Conflicts of interest
}
There are no conflicts to declare.

\section*{Acknowledgements}

I would like thank Stefan Glöggler for insightful discussions and helpful suggestions. 
This project was supported by the Cancer Prevention and Research Institute of Texas (CPRIT) under Grant No. RR240015.

\renewcommand\refname{References}

%%%REFERENCES%%%
\bibliography{rsc} 
\bibliographystyle{rsc} 

\end{document}